\documentclass{sf2a-conf2013}
\usepackage{graphicx}
\usepackage{hyperref}
\usepackage[]{natbib}  
\usepackage{epstopdf}

\def\BibTeX{{\rm B\kern-.05em{\sc i\kern-.025em b}\kern-.08em
    T\kern-.1667em\lower.7ex\hbox{E}\kern-.125emX}}
\bibpunct{(}{)}{;}{a}{}{,}  
\newcommand{\kms}{{\mathrm{km~s^{-1}}}}

\usepackage{bigints}

\begin{document}

\TitreGlobal{SF2A 2013}


\title{Astrophysical false positives in exoplanet transit surveys:\\ why do we need bright stars ?}

\runningtitle{Astrophysical false positive in exoplanet-transit surveys}

\author{A. Santerne}\address{Centro de Astrof\'isica, Universidade do Porto, Rua das Estrelas, 4150-762 Porto, Portugal}


\author{R.~F. D\'iaz}\address{Aix Marseille Universit\'e, CNRS, LAM (Laboratoire d'Astrophysique de Marseille) UMR 7326, 13388, Marseille, France}

\author{J.-M. Almenara$^{2}$}

\author{A. Lethuillier$^{2}$}

\author{M. Deleuil$^{2}$}

\author{C.~Moutou$^{2,}$}\address{CNRS, Canada-France-Hawaii Telescope Corporation, 65-1238 Mamalahoa Hwy., Kamuela, HI 96743, USA}




\setcounter{page}{237}


\maketitle


\begin{abstract}
Astrophysical false positives that mimic planetary transit are one of the main limitation to exoplanet transit surveys. In this proceeding, we review the issue of the false positive in transit survey and the possible complementary observations to constrain their presence. We also review the false-positive rate of both \textit{Kepler} and \textit{CoRoT} missions and present the basics of the planet-validation technique. Finally, we discuss the interest of observing bright stars, as \textit{PLATO~2.0} and \textit{TESS} will do, in the context of the false positives. According to simulations with the Besan\c con galactic model, we find that \textit{PLATO~2.0} is expected to have less background false positives than \textit{Kepler}, and thus an even lower false-positive rate.
\end{abstract}



\section{Astrophysical false positives in transit surveys}

Transiting exoplanets are the only planet for which it is possible to measure independently their mass and their radius. From these measurements, it is then possible to determine their bulk density and to model their internal structure. Since they pass in front or behind their host star, it is also possible to probe their atmosphere composition through transmission or emission spectroscopy. Therefore, transiting exoplanets strongly constrain theories of planet formation, migration and evolution \citep[e.g.][]{2009A&A...501.1161M, 2012A&A...547A.112M}.\\

\begin{figure}[ht!]
 \centering
 \includegraphics[width=0.15\textwidth,clip]{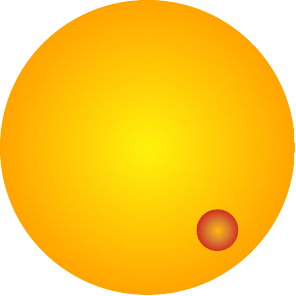}%
 \hspace{0.2cm}
 \includegraphics[width=0.15\textwidth,clip]{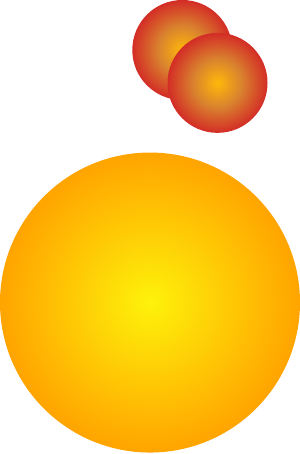}%
 \hspace{0.2cm}
 \includegraphics[width=0.15\textwidth,clip]{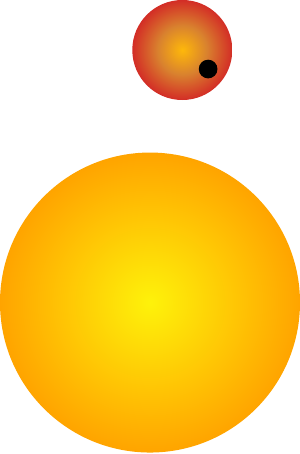}%
  \caption{Sketches of the main false-positive scenarios occurring in transit surveys. From left to right: undiluted eclipsing binary (e.g. eclipsing low-mass star); background eclipsing binary or eclipsing binary in triple system; background transiting planet or companion transiting planet}
  \label{santerne1:fig1}
\end{figure}

Many photometric-transit surveys are searching for new transiting exoplanets, from the ground with e.g. SuperWASP, HATNet, etc\dots \citep{2007MNRAS.375..951C,2007ApJ...656..552B} and from space with \textit{CoRoT} \citep{2006cosp...36.3749B} and \textit{Kepler} \citep{2009Sci...325..709B}. However, searching for new transiting exoplanets is not an easy task. Many configurations of diluted eclipsing binaries or diluted transiting planet might mimic the photometric transit of an exoplanet \citep[and Fig. \ref{santerne1:fig1}]{2012Natur.492...48C, 2013arXiv1307.2003S}. If they are not rigorously identified, those fake exoplanets (so-called ``false positives'') might bias the distributions of planets used to constrain theories of planet formation, migration and evolution. More importantly, those false positives might lead planet theorists to wrong conclusions (C. Mordasini, Planet Validation Workshop).\\

To establish a new transiting exoplanet, one should first check that the observed photometric signal is due to a planet or to a false-positive scenario. For that, different kind of complementary observations can be used:
\begin{itemize}
\item Ground-based high-resolution photometry \citep{2009A&A...506..343D} or centroid measurement \citep{2013arXiv1303.0052B} to reject background eclipsing binary contaminating the target's PSF \citep{2009A&A...506..337A}. Adaptive optics images \citep{2012AJ....144...42A} or speckle observations \citep{2011AJ....142...19H} can also be used to constrain, closer to the star, the presence of a contaminant.
\item Infrared photometry to constrain the presence of a contaminating star with a different color than the target \citep{2012ApJ...745...81F}.
\item High-resolution spectroscopy to identify multiple stellar systems \citep[e.g.][]{2012A&A...545A..76S}.
\end{itemize}

Precise radial-velocity (RV) observations can be used to measure the mass of the transiting object. If this transiting object has a mass compatible with the planet's mass range, the planet is therefore established \citep[e.g.][]{2011A&A...528A..63S, 2011A&A...536A..70S}.\\

\section{The false-positive probability}
\label{santerne1:FPP}

\begin{figure}[ht!]
 \centering
 \includegraphics[width=0.8\textwidth,clip]{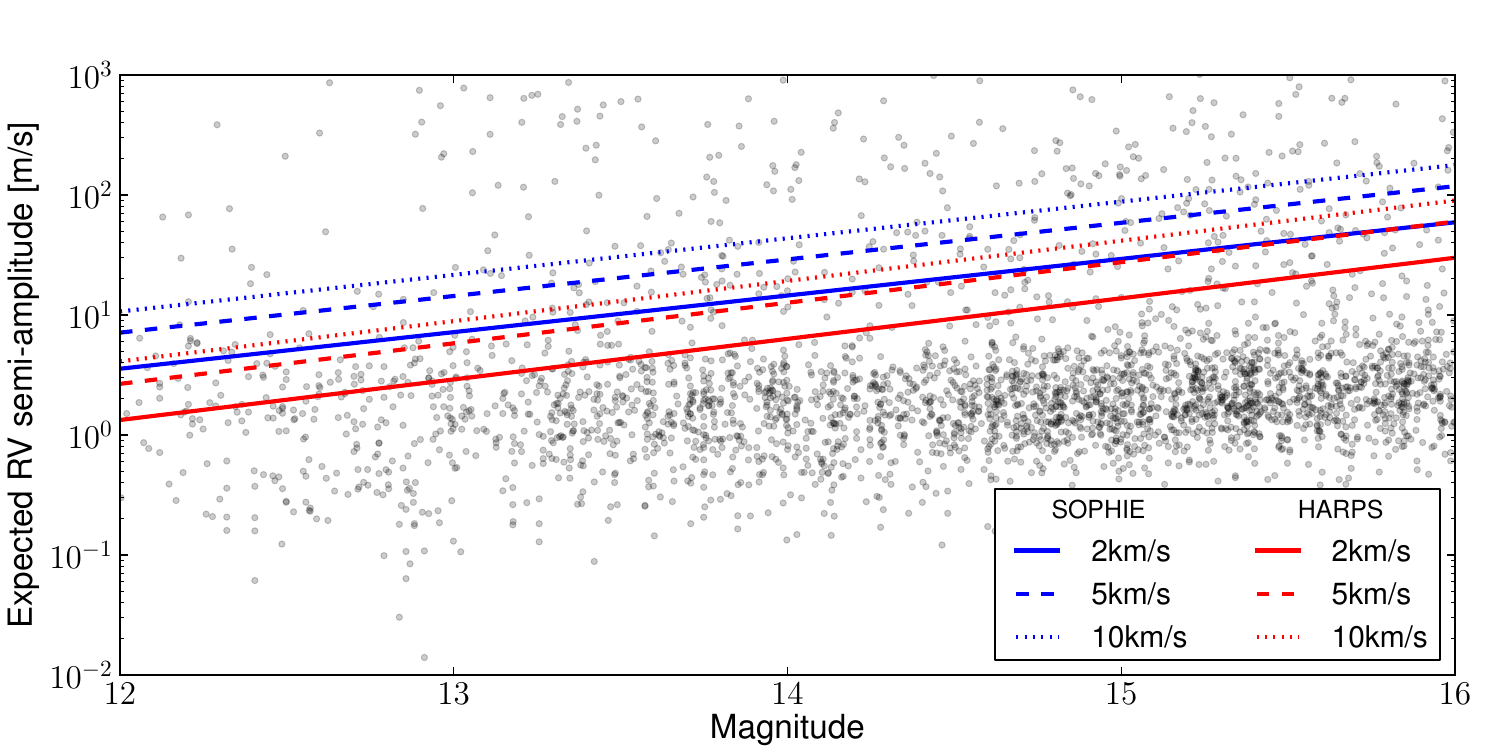}      
  \caption{Expected radial velocity amplitude of the \textit{Kepler} candidates as function of the magnitude of their host star. The mass of the candidates, needed to estimate the amplitude, has been estimated using the estimated radius and densities of solar system objects. The red and blue solid, dashed and dotted lines represent the photon noise limitation in an one-hour exposure time observation with HARPS and SOPHIE (respectively) for a star with a vsini of 2$\kms$, 5$\kms$ and 10$\kms$ (respectively).}
  \label{santerne1:fig2}
\end{figure}

The \textit{CoRoT} and \textit{Kepler} space missions have discovered respectively $\sim$ 600 (Deleuil et al., in prep.) and $\sim$ 3000 exoplanet-candidates \citep{2013ApJS..204...24B} around host stars of magnitude ranging between $\sim$ 10 and 16. Radial velocity follow-up observations of such faint stars are limited by the photon noise \citep{2011EPJWC..1102001S}. Figure \ref{santerne1:fig2} displays the expected radial velocity semi-amplitude of the \textit{Kepler} candidates (assuming an Earth density for KOIs smaller than 2.5~R$_\oplus$, and the density of Neptune for those larger than 2.5~R$_\oplus$) as function of the magnitude of the host star. The majority of the \textit{Kepler} candidates are expected to present a RV signal at the level of a few m.s$^{-1}$ on stars fainter than the 14$^\mathrm{th}$ magnitude. Such precision is below the photon noise of current spectrographs, like SOPHIE and HARPS, in one hour of exposure time \citep[and Fig. \ref{santerne1:fig2}]{2011EPJWC..1102001S}. We therefore anticipated that only a small fraction (5\% to 10\%) of all the \textit{Kepler} candidates can be established as bona-fide planet by measuring the Doppler reflex motion of the host star. For the \textit{CoRoT} candidates, only 5\% of the candidates have been established as planet and another $\sim$ 5\% of the candidates are not resolved pending future observations with improved capabilities (e.g. with ESPRESSO on the ESO--VLT). \\

Identifying the false positive detections is a crucial task. Since only a small fraction of the \textit{Kepler} and \textit{CoRoT} candidates can be established as bona-wide planets, one alternative solution is to estimate the false positive probability (FPP) of the candidates. If the FPP is very low, planets statistics used to constrain theories might be done based on the candidates. Unfortunately, even if the FPP is low in average, there might exists some regions of the candidates parameters space which are more affected by the false positives \citep[the FPP is not expected to be a constant value over all the candidates' parameter space;][]{2012A&A...545A..76S, 2013ApJ...766...81F}. If the theories try to reproduce high-FPP regions of the parameters space, it might incorrectly constrains the theories.\\

The FPP of the \textit{CoRoT} mission has been estimated to be around 80\% (Deleuil et al., in prep.). The FPP of the \textit{Kepler} mission has been much more discussed. First, \citet{2011ApJ...738..170M} found a low-value of the FPP, in average lower than 5\% and lower than 10\% for 90\% of the \textit{Kepler} candidates. By observing with the SOPHIE spectrograph at Observatoire de Haute-Provence, \citet{2012A&A...545A..76S} found a much higher FPP ($34.8\% \pm 6.5\%$) for the giant close-in \textit{Kepler} candidates. This observational value is not compatible with the \citet{2011ApJ...738..170M} estimation. Later, \citet{2013ApJ...766...81F} re-estimate the overall \textit{Kepler} FPP to be $9.4\% \pm0.9\%$ by modeling the efficiency of the \textit{Kepler} pipeline to detect planets. This last value, is compatible with the measurement from \citet{2012A&A...545A..76S}, in the giant close-in regime. Finally, \citet{2013arXiv1307.2003S} studied the occurrence of eclipsing binaries for which only the secondary eclipse is seen from the Earth (because of some fine tuning of the orbital eccentricity, argument of periastron and orbital inclination) and re-evaluate the global \textit{Kepler} FPP to $11.3\%Ê\pmÊ1.1\%$.\\

The difference between \textit{Kepler} and \textit{CoRoT} FPP can be explained by the fact that \textit{Kepler} is much more efficient to reject background eclipsing binary and background transiting planet than \textit{CoRoT}, thanks to the measurement of the centroid during the transit. This can also be explained by the fact that the \textit{Kepler} telescope observed at higher latitude in the galactic plane than \textit{CoRoT}, where the stellar background is slightly less dense.\\

The \textit{Kepler} FPP of $11.3\%Ê\pmÊ1.1\%$ is still too high to allow statistical analysis of the transiting planets based on the \textit{Kepler} candidates neglecting the false positives. About 300 of these candidates are thus expected to be impostors (This number corresponds to $\sim$ 1/3 of all the planets discovered since 18 years). Moreover, these impostors can be more common in some regions of the parameter space, as for example, in the giant candidate regime \citep{2012A&A...545A..76S}. According to the recent study of \citet{2013ApJ...766...81F}, most of the \textit{Kepler} false positives are produced by  background eclipsing binaries and planet transiting a star physically bound with the target.

\section{The planet-validation technique}
\label{santerne1:PVT}

To establish the planetary nature of a candidate, another alternative solution is the so-called planet-validation technique \citep{2011AAS...21811206T}. It consists in computing the probability of the planet scenario against an exhaustive set of false-positive scenarios. If the planet scenario is significantly the highest-probable scenario, thus the planet is considered as validated. Such model comparison can only be done in the Bayesian framework in which hypothesis have a probability (this is not the case in the frequentist approach). Basically, the odds ratio between each pair of scenarios is computed as following:
\begin{equation}
\mathcal{O}_{ij} = \frac{p\left(H_{i}|D,I\right)}{p\left(H_{j}|D,I\right)} = \frac{p\left(H_{i}|I\right)}{p\left(H_{j}|I\right)}\cdot\frac{p\left(D|H_{i},I\right)}{p\left(D|H_{j},I\right)} = \frac{p\left(H_{i}|I\right)}{p\left(H_{j}|I\right)}\cdot\frac{\bigints_{\vec{\theta_{i}}}{p\left(\vec{\theta_{i}}\Big\vert H_{i},I\right)\cdot p\left(D\Big\vert\vec{\theta_{i}},H_{i}, I\right)d\vec{\theta_{i}}}}{\bigints_{\vec{\theta_{j}}}{p\left(\vec{\theta_{j}}\Big\vert H_{j},I\right)\cdot p\left(D\Big\vert\vec{\theta_{j}},H_{j}, I\right)d\vec{\theta_{j}}}}\, ,
\end{equation}
where $H_{i}$ is the hypothesis $i$ (e.g. transiting planet or background eclipsing binary, etc\dots), $D$ is the available data, $I$ the \textit{a priori} information and $\vec{\theta_{i}}$ is the parameter space of the model relative to the hypothesis $i$. To compute this equation, it is needed to compute, first, the hypothesis prior factor (first part of the equation), and then, the Bayes' factor (second part of the equation) which is the ratio between the two hypothesis \textit{posterior} distributions marginalized over the whole parameter space. While the Bayes' factor is estimated from the data using tools such as \texttt{PASTIS} \citep{pastispaper}, the first term of this equation required to know the probability of the two considered hypothesis. When validating a planet, this hypothesis prior factor is the ratio between, e.g., the probability that a given star host a planet, over the probability that a given star is aligned by chance with an eclipsing binary.\\

To compute these \textit{a priori} hypothesis probability, the occurrence of planets and binaries as well as the background stellar density are needed. The occurrence rate of planets and binaries have been estimated based on results from dedicated surveys \citep{2010Sci...330..653H, 2011arXiv1109.2497M, 2013ApJ...766...81F, 2003A&A...397..159H, 2010ApJS..190....1R}. The background density can be estimated using galactic star-count models like TRILEGAL \citep{2005A&A...436..895G} or the Besan\c con Galactic Model \citep{2003A&A...409..523R}. Roughly, a target of magnitude $m_{v_{t}}$ which present a transit of depth $\delta_{t}$ might be mimicked by a background eclipsing stellar contaminant of magnitude $m_{v_{c}}$ and depth $\delta_{c}$ following the equation:

\begin{equation}
m_{v_{c}} - m_{v_{t}}= 2.5 \log \left(\frac{\delta_{c}}{\delta_{t}}\right)
\end{equation}
Therefore, an equal mass eclipsing binary with depth $\delta_{c}=50$\% might mimic a 50ppm-depth transit on a star 10 magnitude\footnote{in the same bandpass} brighter. Since \textit{CoRoT} and \textit{Kepler} targeted stars with magnitude up to 16, the population of potential false positives are stars up to magnitude 26. 

\section{Towards brighter stars with \textit{TESS} and \textit{PLATO~2.0} space missions}

Observing stars much brighter than \textit{CoRoT} and \textit{Kepler} targets, such as those in the scope of \textit{TESS} and \textit{PLATO~2.0} have a main interest for transiting exoplanet characterization: radial velocity follow-up will be much more efficient, being limited only by the instrumental precision of spectrographs. With the next-generation spectrographs like ESPRESSO (ESO -- VLT), it will be possible to characterize the mass of planets down to Earth-like planets in the habitable zone. Therefore, it will be possible to constrain the bulk density of \textit{TESS} and \textit{PLATO~2.0} planets with an unprecedented accuracy. In the case of \textit{PLATO~2.0}, stellar mass, radius and age of planet hosts will be determined precisely thanks to simultaneous asteroseismology, improving even more the accuracy of planet's physical parameters. Many other scientific interests (in planetology as well as stellar physics) will be conducted by targeting bright stars and are discussed in \citet{platopaper} in the context of the \textit{PLATO~2.0} mission.\\

\begin{figure}[ht!]
 \centering
 \includegraphics[width=0.45\textwidth,clip]{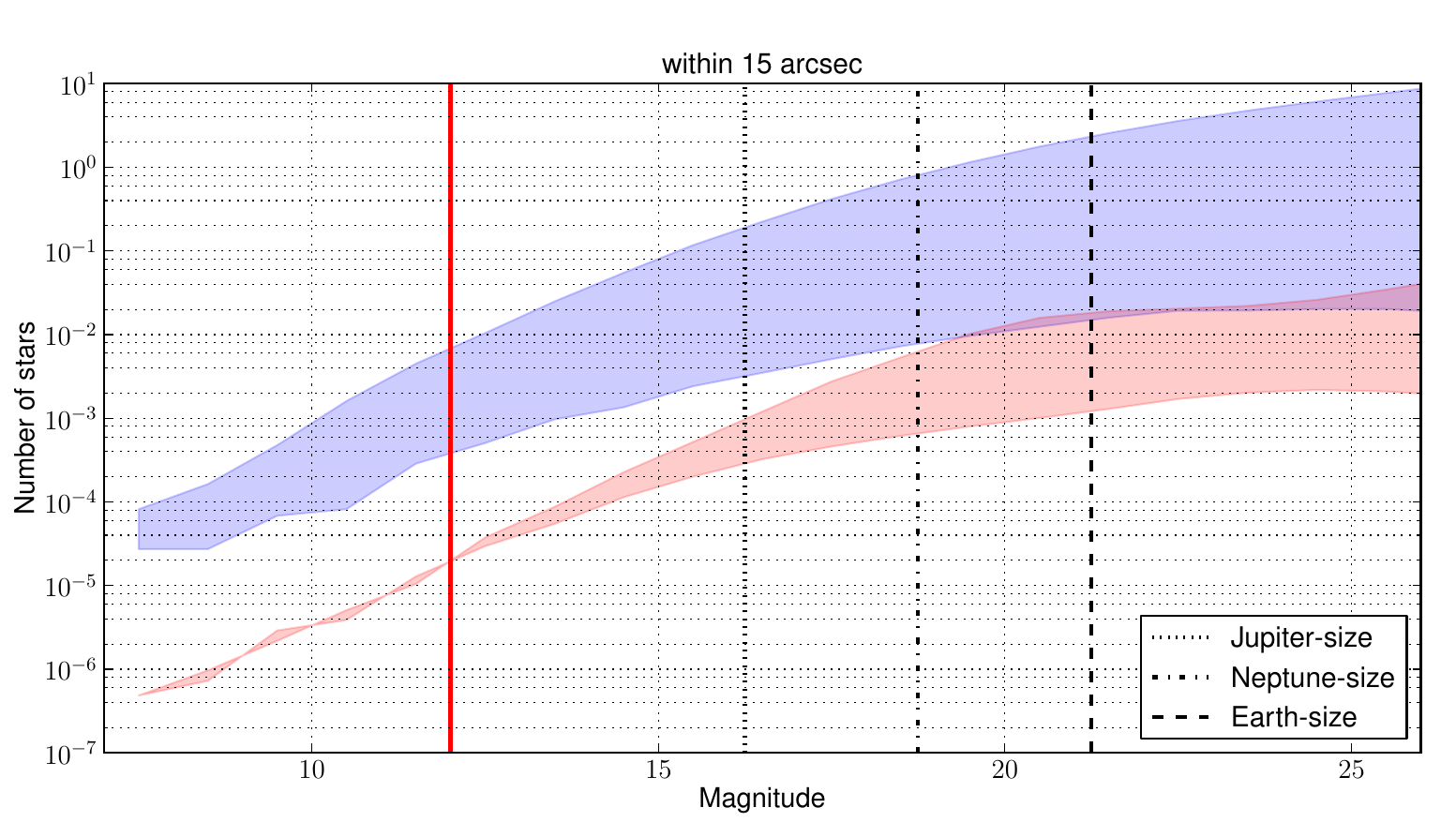}%
 \includegraphics[width=0.45\textwidth,clip]{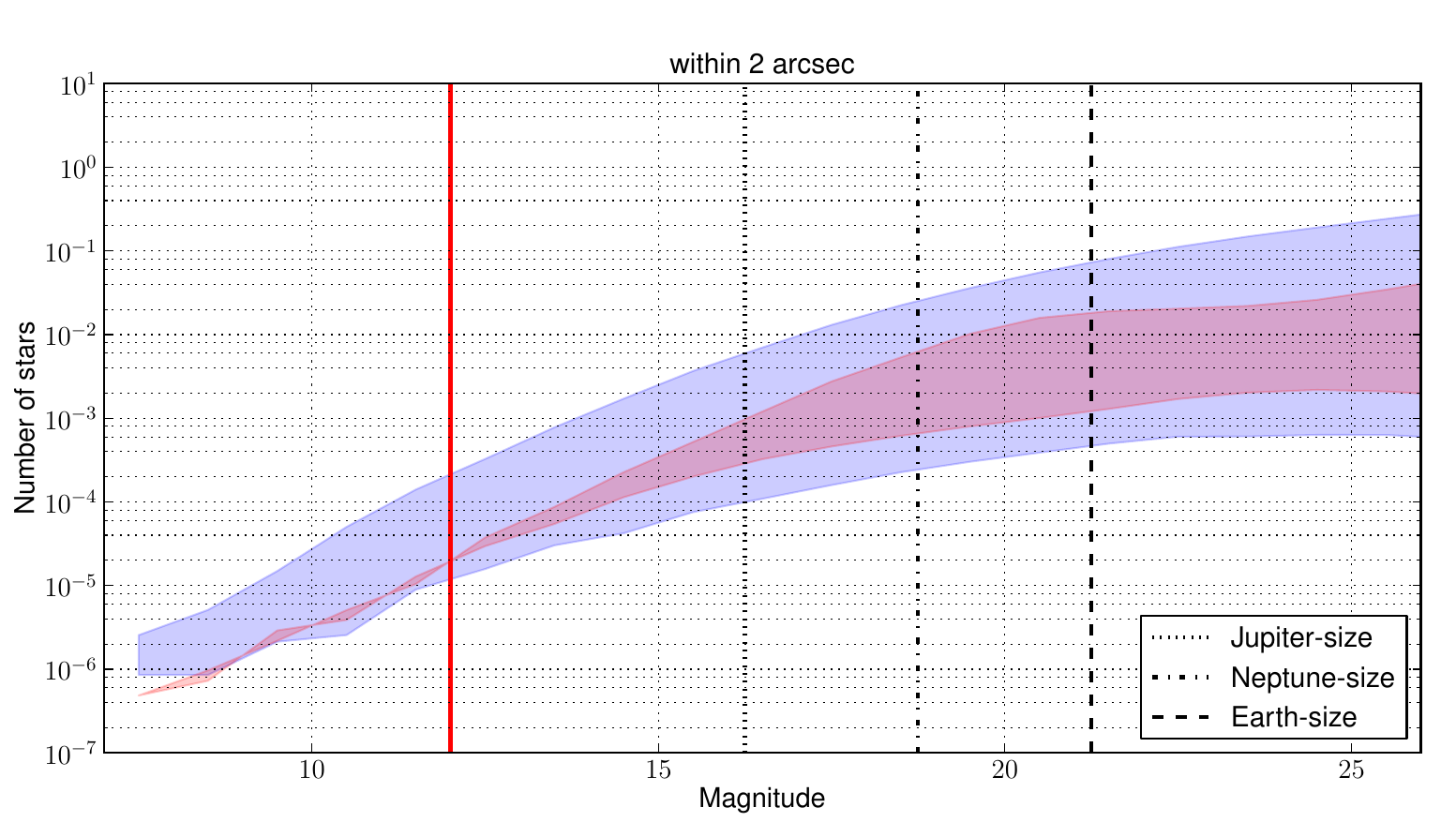}%
  \caption{Background stellar density for \textit{Kepler} field (red area) and \textit{PLATO~2.0} northern field (blue area). The vertical red line indicates a magnitude 12 star and the dotted, dot-dashed and dashed lines indicate the maximum magnitude of a equal-mass eclipsing binary to mimic a Jupiter-size planet, a Neptune-size planet and an Earth-size planet (respectively). \textbf{Left} Background stellar density within a radius of 15 arcsec for \textit{PLATO~2.0} and 2 arcsec for \textit{Kepler}. \textbf{Right:} Background stellar density within a radius of $\sim$ 2 arcsec for both mission.}
  \label{santerne1:fig3}
\end{figure}

Targeting bright stars also have a great interest in terms of astrophysical false positives. Indeed, we might expect that by targeting stars brighter than those observed by \textit{Kepler}, the background stellar density will be much lower, and thus, there will be much lower background false positives (background eclipsing binaries and background transiting planets). This is true only if the instrument's PSF size and the stellar density of the various fields are similar. To qualitatively compare the two effects, we generated field population using the Besan\c con galactic model within 1 deg$^{2}$ up to magnitude R=27 for extreme coordinates in the \textit{Kepler} fields (from b = 5.6$^{\circ}$ -- l = 75.7$^{\circ}$ to b = 20.9$^{\circ}$ -- l = 76.5$^{\circ}$) and in the preliminary-defined northern long run of the \textit{PLATO~2.0} mission (from b = 0$^{\circ}$ -- l = 68.5$^{\circ}$ to b = 65$^{\circ}$ -- l = 40$^{\circ}$). We then extrapolated the star count provided by the Besan\c con galactic model to the exclusion radius of \textit{Kepler} and \textit{PLATO~2.0} (See Fig. \ref{santerne1:fig3}). For \textit{Kepler}, we considered that \textit{Kepler} is able to discard background eclipsing binaries located in a different pixel than the target, hence with an exclusion radius of 2 arcsec\footnote{the pixels of the \textit{Kepler} telescope are 3.96 arcsec large} (See Fig. \ref{santerne1:fig3}). Typically, considering a \textit{Kepler} target of magnitude 16, there is between 0.04 and 0.5 background star within a radius of 2 arcsec that might mimic a planetary transit down to an Earth-size planet. If we assume than \textit{PLATO~2.0} will have the same efficiency than \textit{Kepler} to discard background eclipsing binaries, but with pixels of 15 arcsec \citep{platopaper}, there will be between 0.04 and 3.6 stars in the background of a 12$^{\mathrm{th}}$ magnitude target that might mimic an Earth-size planet. These numbers are much larger than for the \textit{Kepler} mission. Fortunately, \textit{PLATO~2.0} will observe stars with 8 up to 32 telescopes \citep{platopaper}. Assuming that the centroid precision follows the square root of the number of telescope, we find that the maximum number of background stars aligned by chance with \textit{PLATO~2.0} targets of magnitude 12 mimicking down to an Earth-size planet are 0.45 and 0.11 star for 8 and 32 telescopes (respectively). This means that we might expect less background false positives in the \textit{PLATO~2.0} mission compared with \textit{Kepler}, and thus, a lower false-positive rate for \textit{PLATO~2.0} (assuming that the rate of bounded false positives is similar for \textit{Kepler} than \textit{PLATO~2.0}). Since \textit{TESS} will observe stars with only one telescope with large pixels, we expect a large number of background false positives, at least for targets close to the galactic plane.\\

We stress that these results are a rough estimation of the false-positive probability of the \textit{PLATO~2.0} mission, in comparison with \textit{Kepler}. A more rigorous analysis of the expected false-positive probability of this mission will be performed as soon as the \textit{PLATO~2.0} fields and list of targets are defined.

\section{Conclusions and discussion}

Astrophysical false positives are a classical nuisance of exoplanet transit surveys. Neglecting them might lead exoplanet theorists to wrong conclusion. It is therefore crucial to account for them in statistical analysis of transit candidates. For that, the best option is to establish all the candidates by measuring their mass using dedicated Doppler observations. Unfortunately, the \textit{CoRoT} and \textit{Kepler} targets are too faint to allow the characterization of the smallest candidates. The false-positive rate is the key value to perform statistical analysis of the candidates to derive planet properties. For the \textit{CoRoT} mission, the false-positive rate is about 80\% while the \textit{Kepler} false-positive rate has been quite discussed. The latest estimation find a value of $11.3\%Ê\pmÊ1.1\%$ \citep{2013arXiv1307.2003S}. Another possibility is to validate statistically all candidates using tools such as \texttt{PASTIS} \citep{pastispaper}.\\

Next-generation transit space missions (namely \textit{TESS} and \textit{PLATO~2.0}) will target much brighter stars than \textit{CoRoT} and \textit{Kepler}. First, the Doppler observations will be much more efficient to characterize small planets (especially with new spectrographs like ESPRESSO) and then, we might expected to have less false positives in the scope of \textit{PLATO~2.0} than \textit{Kepler} since the background of bright targets is less dense in potential false positives than for faint targets. We therefore anticipate a lower false-positive rate for \textit{PLATO~2.0} than for \textit{Kepler}. On the other hand, \textit{TESS} will observe bright targets with large pixels within which a significant amount of potential false positives might reside. However, this will strongly depends on the galactic latitude of the target. A more detailed study of the expected false-positive probability of both missions can be performed as soon as the target list is defined.

\begin{acknowledgements}
AS acknowledges the support by the European Research Council/European Community under the FP7 through Starting Grant agreement number 239953. RFD is supported by CNES. AS also acknowledges the administrative council of SF2A for providing him a grant to attend this meeting.
\end{acknowledgements}


\bibliographystyle{aa}  
\bibliography{sf2a-template} 

\end{document}